\documentclass[%
aps,
 pra,%
 amsmath,amssymb,
 reprint,%
 superscriptaddress,
]{revtex4-1}

\usepackage{dcolumn}
\usepackage{graphicx}
\graphicspath{{figure/}}
\usepackage{lineno}
\usepackage{bm} %
\usepackage[pdftex,colorlinks=true,bookmarks=false,citecolor=blue,urlcolor=blue]{hyperref}
\usepackage{upgreek}
\usepackage{xcolor}
\usepackage{titlesec} 

\begin{document}

\title{\vspace{-15mm}\fontsize{19pt}{10pt}\selectfont\textbf{A self-starting bi-chromatic LiNbO$_3$ soliton microcomb}} 

\author{Yang He}
\thanks{These two authors contributed equally.}
\affiliation{Department of Electrical and Computer Engineering, University of Rochester, Rochester, NY 14627}

\author{Qi-Fan Yang}
\thanks{These two authors contributed equally.}
\affiliation{T.J. Watson Laboratory of Applied Physics, California Institute of Technology, Pasadena, California 91125, USA}

\author{Jingwei Ling}
\affiliation{Institute of Optics, University of Rochester, Rochester, NY 14627}

\author{Rui Luo}
\affiliation{Institute of Optics, University of Rochester, Rochester, NY 14627}

\author{Hanxiao Liang}
\affiliation{Department of Electrical and Computer Engineering, University of Rochester, Rochester, NY 14627}

\author{Mingxiao Li}
\affiliation{Department of Electrical and Computer Engineering, University of Rochester, Rochester, NY 14627}

\author{Boqiang Shen}
\affiliation{T.J. Watson Laboratory of Applied Physics, California Institute of Technology, Pasadena, California 91125, USA}

\author{Heming Wang}
\affiliation{T.J. Watson Laboratory of Applied Physics, California Institute of Technology, Pasadena, California 91125, USA}

\author{Kerry Vahala}
\affiliation{T.J. Watson Laboratory of Applied Physics, California Institute of Technology, Pasadena, California 91125, USA}

\author{Qiang Lin}
\email[Electronic mail: ]{qiang.lin@rochester.edu}
\affiliation{Department of Electrical and Computer Engineering, University of Rochester, Rochester, NY 14627}
\affiliation{Institute of Optics, University of Rochester, Rochester, NY 14627}

\begin{abstract}

For its many useful properties, including second and third-order optical nonlinearity as well as electro-optic control, lithium niobate is considered an important potential microcomb material. Here, a soliton microcomb is demonstrated in a monolithic high-Q lithium niobate resonator. Besides the demonstration of soltion mode locking, the photorefractive effect enables mode locking to self-start and soliton switching to occur bi-directionally. Second-harmonic generation of the soliton spectrum is also observed, an essential step for comb self-referencing. The Raman shock time constant of lithium niobate is also determined by measurement of soliton self-frequency-shift.  Besides the considerable technical simplification provided by a self-starting soliton system, these demonstrations, together with the electro-optic and piezoelectric properties of lithium niobate, open the door to a multi-functional microcomb providing f-2f generation and fast electrical control of optical frequency and repetition rate, all of which are critical in applications including time keeping, frequency synthesis/division, spectroscopy and signal generation.

\end{abstract}

\maketitle 
On-chip generation of optical frequency combs via the Kerr nonlinearity has attracted significant interest in recent years \cite{kippenbergreview2018} and the use of these devices for comb systems on a chip is being studied across a wide range of applications including spectroscopy \cite{Vahala162,dutt2018}, communications \cite{Koos17}, ranging \cite{Vahala18, Koos18}, frequency synthesis \cite{Diddams18}, astrocombs \cite{obrzud2017,suhastrocomb2018} and optical clocks \cite{Newman2018}. The recent realization of stable mode locking of Kerr microcombs \cite{Kippenberg14}, wherein the comb lines are mutually phase locked to form a stable soliton pulse train in time, is crucial for all of these applications. First observed in optical fiber systems \cite{Leo2010} these coherently pumped solitons \cite{Wabnitz1993} have now been demonstrated in microresonators made from magnesium fluoride \cite{Kippenberg14, Maleki15, Kippenberg172}, silica \cite{Vahala15, Diddams17}, silicon \cite{Gaeta162}, and silicon nitride \cite{Kippenberg162, Gaeta16, Wang2016, Kartik17, Kippenberg172}. 

In this work, we report generation of Kerr solitons in high-Q lithium niobate resonators. Lithium niobate (LN) (both in bulk form and when fabricated into waveguides and resonators) is of considerable interest for its versatile properties including the electro-optic as well as  second and third-order nonlinear optical effects  \cite{Reano14, Fathpour14, Loncar142, Bower16, Luo17, Liang17, Buse17, Loncar18, Shayan18, Loncar182}. And the combining of these properties with stable soliton mode locking significantly amplifies the functionality of soliton microcombs. Moreover, another property of LN that has attracted somewhat less attention (the photorefractive effect \cite{GunterBook}) is shown to enable a self-starting soliton microcomb as well as bi-directional switching control of multi-soliton states. Besides this unusual and useful property of the LN microresonator system, intracavity second harmonic generation of the soliton microcomb is demonstrated using LN's  quadratic nonlinearity, thereby providing a path for f-to-2f self-referencing without the need for external frequency doublers. Finally, the self-frequency shift of soliton pulses is measured versus soliton bandwidth in order to determine the Raman shock time constant of LN. 
 
\medskip

\noindent{\bf Photorefractive induced bistability:} The processes for triggering the soliton mode-locked state and for switching between states having different soliton number are complicated by the presence of a well known thermo-optic effect in high-Q resonators \cite{Carmon2004}. Because solitons significantly shift the cavity resonance towards lower frequencies (via the Kerr effect), they must be optically pumped at a frequency that is red-detuned relative to the passive cavity resonance \cite{Kippenberg14, Coen13}. This detuning regime, however, is difficult to access directly because it is inherently unstable on account of the thermo-optic nonlinearity \cite{Carmon2004}. Accessing the soliton state has therefore prompted introduction of several techniques for power and frequency control so as to trigger and stabilize solitons \cite{Kippenberg14, Vahala163, Gaeta16, Kippenberg172}. Also, methods for deterministic control of soliton number through backward tuning have been developed as a direct result of the thermo-optic effect \cite{guo2016}. This form of tuning control is, however, unidirectional meaning that soliton number can only be decreased from a initial larger value that is itself determined by a stochastic process.   

\begin{figure*}[htbp]
	\centering\includegraphics[width=2\columnwidth]{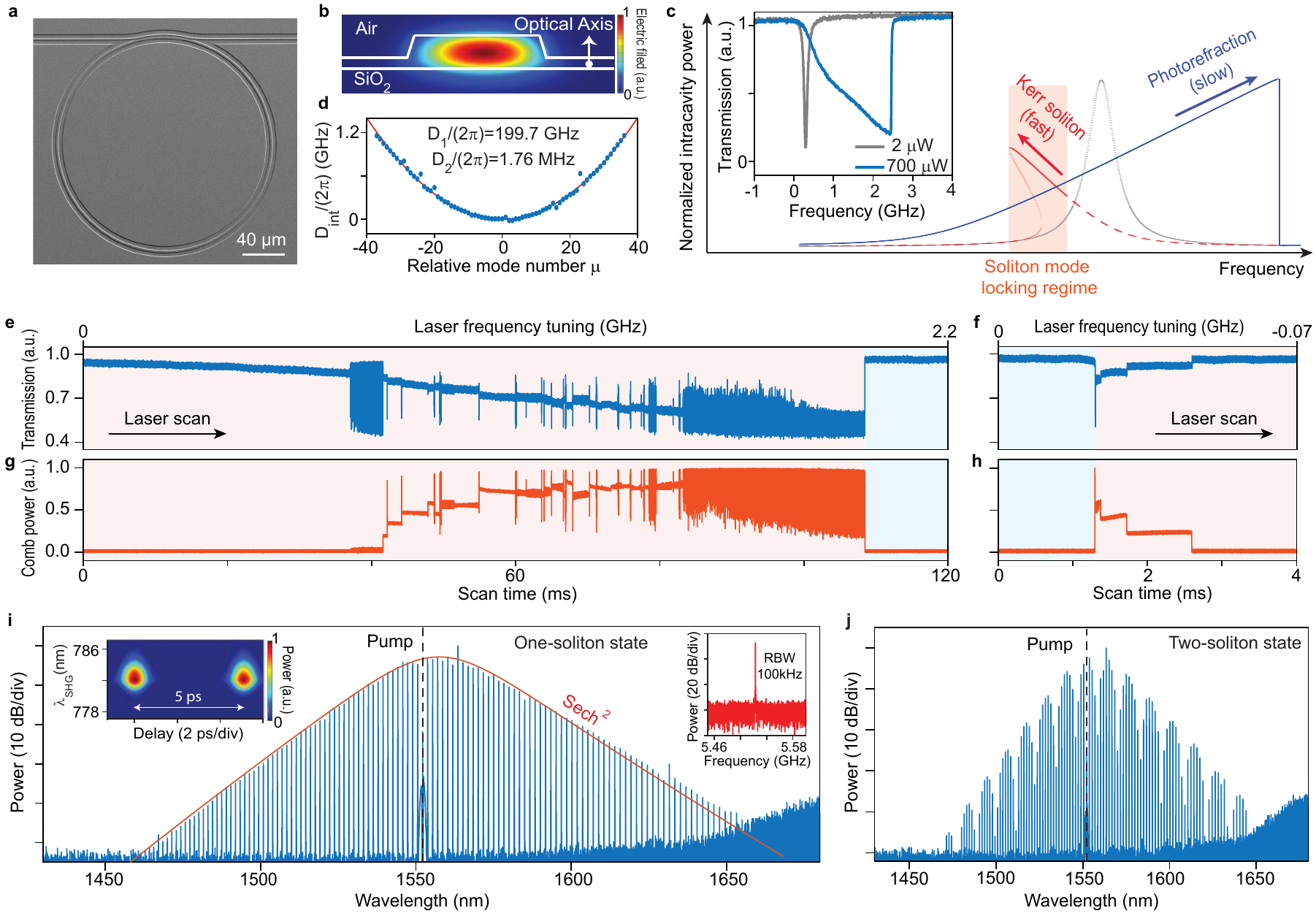}
	\caption{{\bf Lithium niobate microring resonator and mode-locked Kerr solitons.} {\bf a.} Scanning electron microscope image of a lithium niobate microring with a radius of 100~${\rm \mu m}$. {\bf b.} Schematic of the waveguide cross section of the z-cut lithium niobate microring and the simulated optical mode profile of the fundamental quasi-TE cavity mode. {\bf c.} Schematic showing the resonance tuning induced by the optical Kerr effect, which shifts the resonance towards the red (red curve) on a short time scale, and the photorefractive effect, which shifts the resonance towards the blue (blue curve) on a much longer time scale.  As a result, the stable soliton formation regime (shaded region) resides within the laser detuning regime that is stabilized by the photorefractive effect, thereby enabling self-starting soliton mode locking. The gray curve  represents the intrinsic cavity resonance (with a certain photorefractive-induced tuning). The inset is measured power transmission versus pump frequency tuning (red to blue) for a quasi-TE cavity mode at two pump powers. {\bf d.} Measured dispersion $D_{\rm int}$ as a function of the relative mode number ${\rm \mu}$. See Methods for details. {\bf e and f.} Resonator transmission when the pump laser frequency is scanned from red to blue ({\bf e}) and blue to red ({\bf f}) across the pump resonance. The laser power is 33~mW on chip. The shaded red and blue regions correspond to approximate red-detuned and blue-detuned regions, respectively. {\bf g and h.} Comb power versus frequency tuning corresponding to ({\bf e}) and ({\bf f}), respectively. {\bf i.} Optical spectrum of the single-soliton state measured at the first step of the comb power in ({\bf g}). The red curve is a fitting to the theoretical ${\rm sech^2}$ soliton spectral envelope. The left inset shows the corresponding FROG trace. The right inset shows the spectrum of a beat note generated by heterodyning a comb line with a reference CW laser at around 1566.8~nm. {\bf j.} Optical spectrum of the two-soliton state measured at the second step of the comb power ({\bf g}).  In {\bf i} and {\bf j}, the dashed lines indicate the spectral location of the pump mode. } 
	\label{Fig1}
\end{figure*}

This situation is very different in the LN system. Here, the photorefractive effect causes an intensity-dependent decrease of refractive index \cite{GunterBook} (opposite to the thermo-optic effect), and moreover the thermo-optic coefficient of the ordinary polarization in LN is relatively small around room temperature \cite{Rendina05}.  Therefore, if a microresonator is fabricated on a z-cut LN wafer, the quasi-transverse-electric (quasi-TE) cavity modes (Fig.~\ref{Fig1}a,b) will exhibit an optical bistability that stabilizes operation of the pump for red detuned frequencies relative to the cavity resonance.  Fig.~\ref{Fig1}c schematically illustrates the combined Kerr and photorefractive nonlinear hysteresis curves versus pump frequency. The approximate tuning range where solitons form is also indicated. The blue photorefractive tuning curve (provided in the schematic as intracavity optical power) is measured as power transmission versus tuning within the inset of Fig.~\ref{Fig1}c.  There, a cavity resonance is scanned using a tunable laser at two power settings.  At the higher power level, the resonance features a triangular shape versus pump frequency that faces towards higher frequency (opposite to that induced by the thermo-optic effect). This photorefractive-induced behavior stabilizes the laser-cavity detuning when the pump is red-detuned where the soliton is formed (Fig.~\ref{Fig1}c) and enables the soliton mode locking process to self-start. Also, as discussed below, the combination of the relatively slow photorefractive nonlinearity with the fast Kerr effect leads to dynamics that enable bi-directional control of the soliton states.

\begin{figure*}[htpb]
	\centering\includegraphics[width=2\columnwidth]{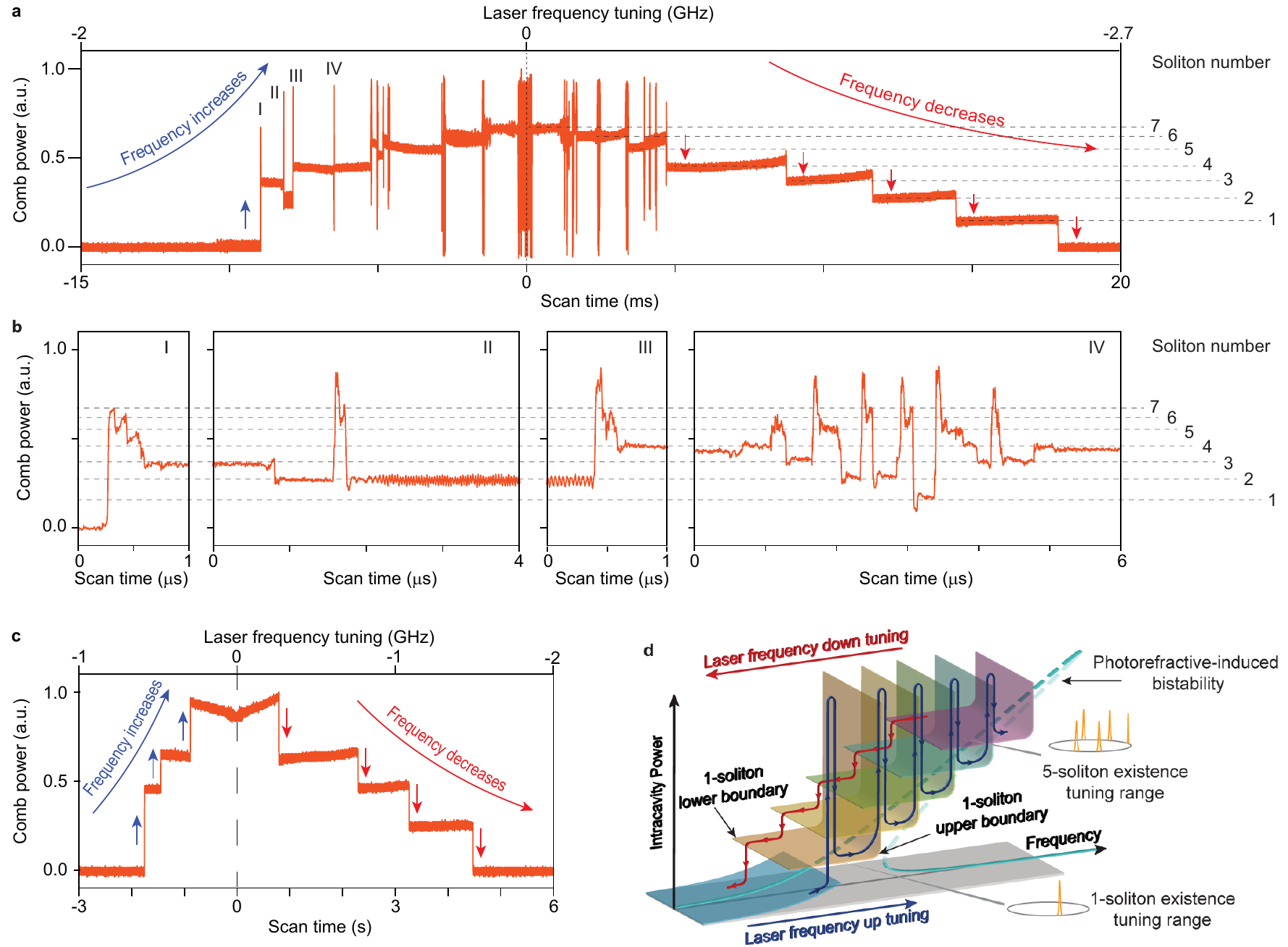}
	\caption{{\bf Bi-directional switching of soliton states.} {\bf a.} Comb power measured as a function of time when the laser frequency is scanned forward and backward across a few soliton steps. The dashed lines give the number of solitons. {\bf b.} Temporally resolved waveforms of the power spikes indicated as I-IV in {\bf a}. The dashed lines indicate the number of solitons corresponding to those in {\bf a}. {\bf c.} Comb power measured as in {\bf a} but using a slower laser frequency scanning speed.  {\bf d.} Schematic showing the bi-directional switching of soliton states. Each L-shaped colored surface represents the existence frequency tuning range of a soliton state \cite{Kippenberg14}. Different colors of L-shaped surfaces represent distinct soliton number states. The photorefractive effect adiabatically blue shifts the cavity resonance (dashed blue curve in the background) and thus disperses the existence tuning ranges of different soliton states over different frequency ranges. The dark blue curve gives the path of increasing soliton number states when the laser frequency is increased, while the red curve gives the path of decreasing soliton number states when the laser frequency is decreased. } \label{Fig2}
\end{figure*}

\medskip

\noindent{\bf Mode-locked soliton states:} To demonstrate both soliton generation in LN microresonators as well as the self-starting feature, we employ a z-cut LN microresonator (Fig.~\ref{Fig1}a,b) with a loaded optical Q factor of $2.2\times 10^6$, a free-spectral range (FSR) of 199.7~GHz, and an anomalous group-velocity dispersion of $\beta_2 = -0.047~{\rm ps^2/m}$ ($D_2/(2\pi)$ = 1.76~MHz) in the telecom band (Fig.~\ref{Fig1}d). The experimental setup is shown in the Supplementary Information. Pump power of 33~mW is coupled onto the chip. The pump transmission (Fig.~\ref{Fig1}e,f) and comb power (Fig.~\ref{Fig1}g,h) are monitored for pump frequency scanning in both tuning directions. The comb power versus tuning shows discrete steps (Fig.~\ref{Fig1}g,h), a signature of soliton mode locking. Moreover, by stopping the laser scan it is possible to observe single and multi-soliton states. At the first power step in Fig.~\ref{Fig1}g  the output spectrum is observed to exhibit a smooth ${\rm sech^2}$-shaped spectral envelope (Fig.~\ref{Fig1}i) with a 3-dB bandwidth of $\sim$27.9~nm.  The  mode spacing is equal to the FSR, indicating a single-soliton state. At the second power step, a sinusoidal modulation with a period of 12~nm is super-imposed on the ${\rm sech^2}$-shaped optical spectrum (Fig.~\ref{Fig1}j), indicating a two-soliton state. At higher power steps, the spectral modulation becomes more complicated while the optical spectrum still maintains an overall ${\rm sech^2}$ shape (not shown), indicating multi-soliton states. Accessing the soliton regime from the red detuning side as observed here has not been reported before and allows the soliton regime to be entered without the need for triggering techniques \cite{kippenbergreview2018}. On the other hand, when the pump frequency is scanned from blue-detuned frequencies across the resonance the step formation is not observed until the pump frequency enters into the red-detuned regime (Fig.~\ref{Fig1}h), at which point the power steps correspond to the first few steps in Fig.~\ref{Fig1}g. 

\begin{figure*}[htbp]
	\centering\includegraphics[width=2\columnwidth]{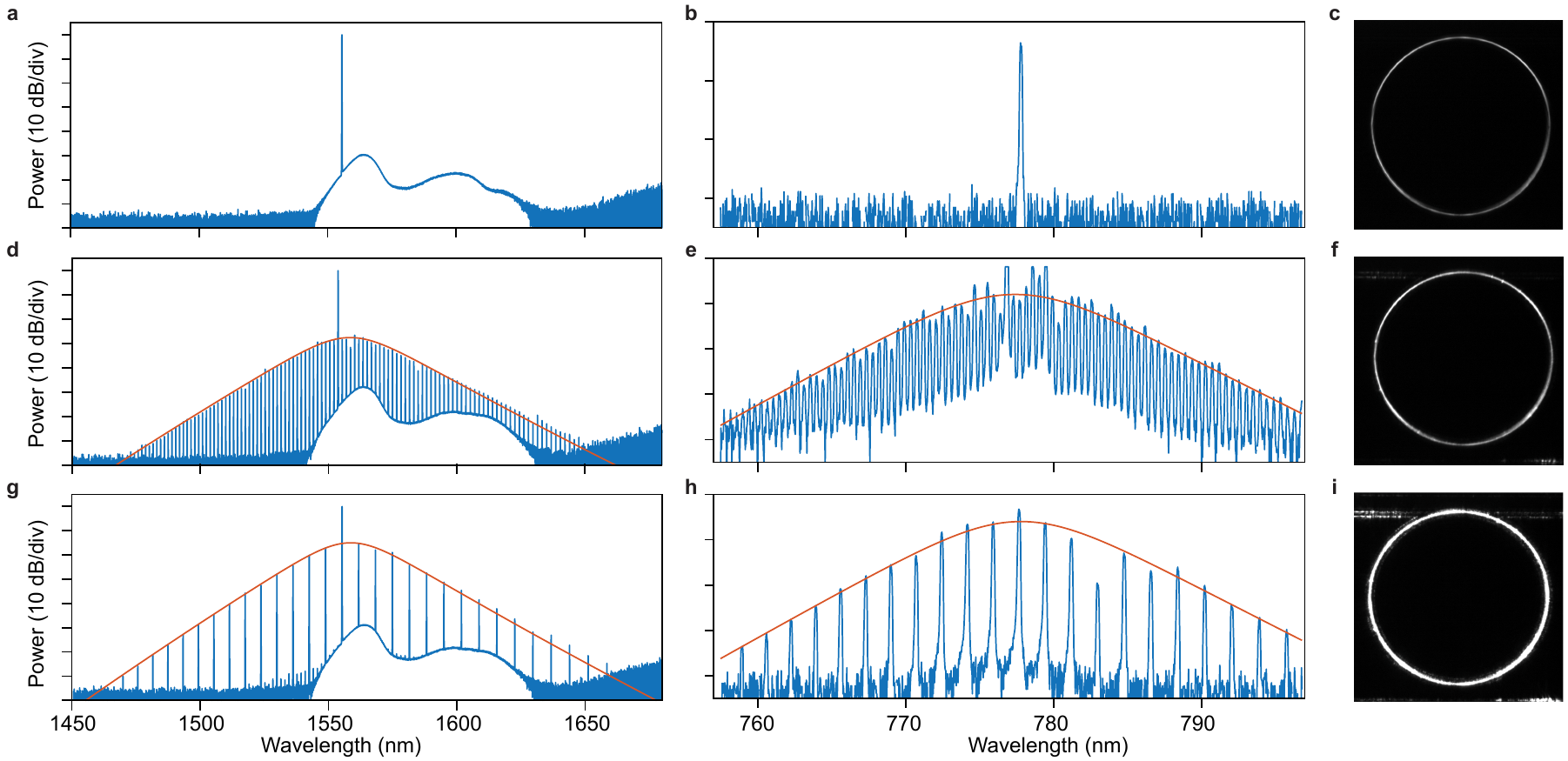}
	\caption{{\bf Second harmonic generation.} {\bf a, d, g.}  Optical spectra in the telecom band: ({\bf a}) before comb generation, ({\bf d}) for a single soliton state , and ({\bf g}) for a soliton crystal state . The pump power is 23~mW on chip and the spectra were recorded with a spectral resolution of 0.05~nm. {\bf b, e, h.} Optical spectra of generated second harmonic corresponding to {\bf a, d}, and {\bf g}, respectively. The spectra were recorded with a resolution of 0.15~nm. In {\bf d, e, g}, and {\bf h}, the red curves are fitted ${\rm sech^2}$ spectral envelopes. {\bf c, f, i.} Optical microscope images showing second-harmonic scattering from the resonator, corresponding to {\bf b, e} and {\bf h}, respectively. } \label{Fig3}
\end{figure*}  

To further verify the single soliton state at the first power step, we characterized the temporal waveform of the output pulses by frequency-resolved optical gating (FROG). As shown in the left inset of Fig.~\ref{Fig1}i, the recorded FROG spectrogram indicates pulse waveforms with a period of 5 picoseconds, corresponding to the round-trip-time of the microresonator. To further characterize the coherence of the comb lines, a tunable external cavity diode laser was heterodyned with a microcomb line. As shown in the right inset of Fig.~\ref{Fig1}i, the beat note exhibits a signal-to-noise ratio greater than 40~dB indicating the high coherence of the comb lines.  Numeric modeling using the Lugiato-Lefever equation \cite{Coen132} was performed to confirm the self-starting mode locking process via the photorefractive effect. Details are provided in the Supplementary Information.


\medskip 

\noindent{\bf Bi-directional switching of soliton states:} Of particular interest is that the stabilization introduced by the photorefractive process enables bi-directional switching between different soliton states, a phenomenon that has so far not been possible in other soliton microcomb systems \cite{kippenbergreview2018}. To demonstrate this phenomenon, the laser frequency was scanned up and down across a few soliton steps for red-detuned operation. As shown in Fig.~\ref{Fig2}a, when the laser frequency increases the comb power climbs up along discrete steps, indicating that the Kerr comb transits from lower number to higher number soliton states. Then, when the laser frequency decreases the comb power steps down discretely, indicating a reverse transitioning of soliton states. Increasing soliton number steps are observed to exhibit richer dynamics compared with the decreasing number process. Specifically, an increase in soliton number is frequently accompanied by a power spike,  while decreasing steps  have no spiking behavior.  Also, temporal resolution of the power spikes reveals a step-like substructure as shown in Fig.~\ref{Fig2}b, I-IV. These spikes become less prominent when the laser scan speed is slowed as in Fig.~\ref{Fig2}c.

The soliton number step-up and step-down process (increasing and decreasing pump frequency) as well as the power spike phenomenon can be understood qualitatively by considering the schematic of microcomb power versus pump frequency tuning  in  Fig.~\ref{Fig2}d. In addition, the power spike and substructure are numerically modeled in the Supplementary Information. In the schematic the laser frequency scan rate is assumed to be slow enough so that at any point in the scan the photorefractive effect is approximately in steady state with respect to the cavity optical power.  The dynamic response of the photorefractive effect  to optical power is complex but it is measured to have a fast component of response that is $\sim 8~{\rm \mu s}$, see Supplementary Information, which sets an approximate time scale for the slow frequency scan in the schematic. In contrast, the ultra-fast Kerr effect responds nearly instantaneously to cavity power changes. Now consider a pump laser scanning towards the cavity resonance from the red detuned side (blue trajectory in Fig.~\ref{Fig2}d).  The detuning of the pump frequency relative to the cavity resonant frequency determines dynamical regimes of the Kerr nonlinear system.  Solitons exist within a specific red detuning range of the pump frequencies relative to the cavity resonance \cite{Kippenberg14}. As the pump tunes, it enters this range, ultimately exits near the resonance and briefly enters a regime where instabilities exist leading to the onset of the power spike.  The accompanying increase in cavity power also causes a slower photorefractive tuning of the resonance away from the pumping frequencies, thereby returning the system into the soliton detuning regime.  The location of the photorefractive-shifted cavity resonance is illustrated by the dashed blue hysteresis curve in Fig.~\ref{Fig2}d. 

Here, begins a relaxation process to a final steady state that occurs on  a time scale set by the photorefractive effect and that typically involves a series of intermediate steps. The overall process occurs within the power spike (illustrated approximately by the blue spikes in Fig.~\ref{Fig2}d).  First, upon re-entering the soliton existence detuning range, a soliton state is acquired with a certain number N of solitons. This occurs very quickly on a time scale set by the cavity lifetime (in this case $\sim$ 1 ns).  The soliton state N, through the corresponding increased cavity power, continues the slower cavity tuning via the photorefractive effect and further pushes the cavity resonant frequency towards higher frequencies and away from the pumping frequency.  This photorefractive-induced tuning can cause the detuning of pumping frequency relative to the cavity resonance to exit the soliton existence range,  thereby forcing the system to acquire another soliton state. This process continues until  the resulting detuning of the pump relative to the photorefractive-shifted cavity resonance is consistent with the soliton number that is inducing the photorefractive cavity shift. Because smaller soliton numbers cause smaller photorefractive shifts, the self-consistent soliton number solution naturally evolves from low soliton number states to larger soliton numbers as the pump approaches the resonance from the red-detuned side.  

Measurements of the transient power spiking process with high temporal resolution are presented in Fig.~\ref{Fig2}b for four power spikes in Fig.~\ref{Fig2}a. The substructure can be readily associated with passage through meta-stable soliton states described above as the system ultimately achieves a self-consistent solution. 
The switching to higher number soliton states can become more regular and ordered if the scanning speed of laser allows the photorefractive effect to stably settle. This is shown in Fig.~\ref{Fig2}c where a slow frequency scan of the laser is applied.  Finally, downward frequency tuning of the laser (red curve in Fig.~\ref{Fig2}d) occurs in a similar manner. However, here, because the system is initially in a soliton state and because tuning occurs across the far edge of the soliton existence boundary (i.e., larger pump-cavity detuning) the system is never within the modulation instability regime so that power spiking does not occur.

\medskip

\noindent{\bf Second harmonic generation of Kerr solitons:} Lithium niobate exhibits a significant quadratic optical nonlinearity, which can enable the frequency up-conversion of the Kerr solitons inside the cavity to produce a bi-chromatic soliton microcomb. This conversion is readily visible in the light scattered from the resonator as captured by a camera  and spectrometer (see Supplementary Information for the experimental setup). As shown in Fig.~\ref{Fig3}a,b, before the soliton microcomb is produced the pump laser launched into the cavity generates only a small amount of up-converted light. In this condition weak second-harmonic emission is visible in the light scattered from the microresonator (Fig.~\ref{Fig3}c). However, when the soliton is formed (Fig.~\ref{Fig3}d), its high peak power enhances the up-conversion process, resulting in a brighter image of the scattered second harmonic light (Fig.~\ref{Fig3}f). Also, as shown in Fig.~\ref{Fig3}e, the spectrum of the second harmonic exhibits an overall ${\rm sech^2}$ shape with a 3-dB bandwidth of about 9~nm (corresponding to $\sim$4.47~THz). The second harmonic spectrum is broader than that of the single soliton in the telecom band ($\sim$25.3~nm in Fig.~\ref{Fig3}d, corresponding to $\sim$3.13~THz). This is expected since the efficiency of second harmonic generation depends on the optical power and thus on the temporal profile of the fundamental Kerr soliton.

Second harmonic generation of a soliton crystal (Fig.~\ref{Fig3}g) is shown in Fig.~\ref{Fig3}h and Fig.~\ref{Fig3}i.  Here, the mode spacing of 6.4~nm, 4 times that of the single-soliton state, implies that four solitons are spaced equally in the cavity (corresponding to a soliton pulse train with a repetition rate of $\sim$800~GHz). The overall power of the soliton crystal is higher inside the cavity resulting in higher power of the generated second harmonic and a considerably brighter image of the scattered light, as shown in Fig.~\ref{Fig3}i. The spectrum of the second harmonic again exhibits a clear ${\rm sech^2}$ shape as shown in Fig.~\ref{Fig3}h. 

Frequency up-conversion of Kerr combs has been explored in aluminum nitride microresonators \cite{Tang2018}. However, so far, up-conversion of Kerr solitons has not been demonstrated. The bi-chromatic soliton microcombs observed here show that the lithium niobate soliton resonators have great potential for this purpose. The coupling waveguide in the current device exhibits very low coupling efficiency for the second harmonic signal as evident in the images of Fig.~\ref{Fig3}, since it was designed for operation in the telecom band. The second harmonic coupling efficiency can be significantly improved with future optimization of the coupling-waveguide design \cite{Tang2018}.  

\begin{figure*}[htbp]
	\centering\includegraphics[width=2\columnwidth]{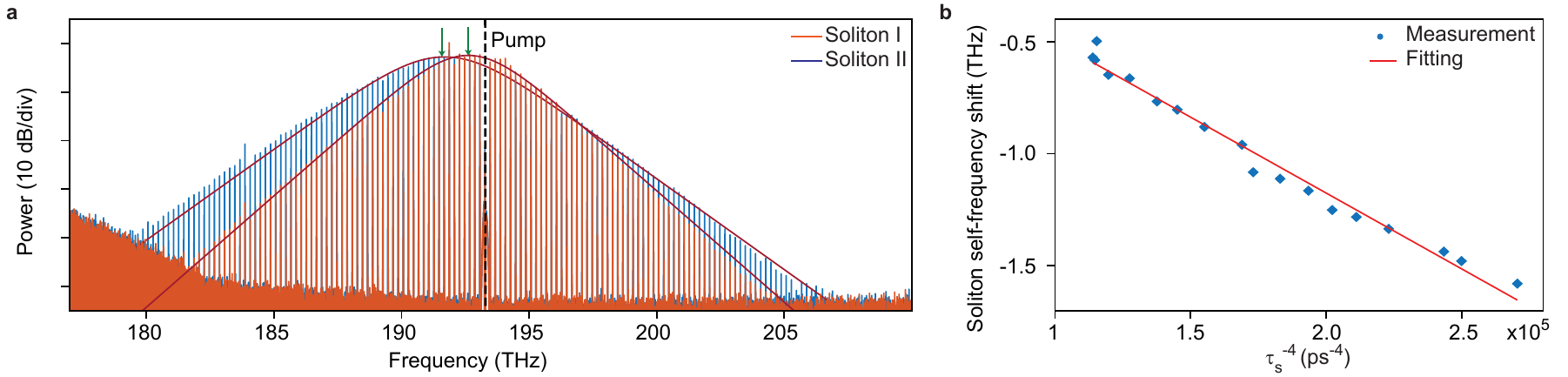}
	\caption{{\bf Raman-induced soliton-self-frequency-shift.} {\bf a,} Optical spectra for two single soliton states having different pumping powers. The pump mode is indicated by the dashed black line and the red curves are the fitted ${\rm sech^2}$ envelopes, whose spectral centers are indicated by the green arrows. {\bf b,} Soliton self-frequency shift as a function of $1/\tau_s^4$ with experimental points shown as blue dots and theoretical fitting line shown in red \cite{Vahala163}.} \label{Fig4}
\end{figure*}


\medskip

\noindent{\bf Self-frequency shift of Kerr solitons:} The optical spectrum of the single-soliton state in Fig.~\ref{Fig1}i is not centered around the pump frequency, but rather is slightly shifted towards lower frequency. The shift has been observed in silica and silicon nitride microresonators \cite{Vahala15, Kippenberg16, Vahala164} and is related to the well known self-frequency shift of solitons induced by the intrapulse Raman scattering \cite{Mollenauer86, Gordon86}.  To characterize this effect, we varied the laser frequency detuning as well as the pump power and monitored the soliton spectrum. Figure~\ref{Fig4}a compares two soliton spectra and shows that a larger spectral bandwidth (corresponding to narrower soliton pulse width) results in a larger self-frequency shift. As shown in Fig.~\ref{Fig4}b, the measured Kerr soliton self-frequency shift exhibits a linear dependence on $1/\tau_s^4$ where $\tau_s$ is soliton pulse width with a magnitude given by $\Omega = \frac{8c\tau_R Q \beta_2}{15 n_0 \omega_0}\frac{1}{\tau_s^4}$  \cite{Vahala164}, where $n_0$, $c$, and $\omega_0$ are the effective refractive index, the velocity of light in vacuum, and the center frequency of the soliton, respectively. $\tau_R$ is the Raman shock time, an important parameter governing the dynamics of ultrashort pulses in Raman active nonlinear media. Although LN is a fairly well studied nonlinear medium, the Raman shock time has never been measured before \cite{Weber66, Ridah97}. The observed soliton self-frequency shift therefore provides an excellent opportunity to characterize this parameter. Fitting the measured linear dependence, we obtain the Raman shock time of congruent LN to be $\tau_R = 6.3$~fs. This value is about three times larger than in fused silica \cite{Vahala164}, indicating the potentially crucial role of the Raman effect on ultashort pulse propagation in LN devices. 

\medskip

\noindent{\bf Discussion:} We have demonstrated a soliton microcomb in a lithium niobate microresonator. Lithium niobate's photorefractive effect enables self-start of soliton mode locking and bi-directional switching of soliton states, and its quadratic nonlinearity enables frequency doubling of the Kerr solitons and soliton crystals. The bi-chromatic soliton microcombs observed here have great potential for realizing direct on-chip f-to-2f signal generation that is essential for comb self referencing. To date, f-to-2f generation must rely on external nonlinear media for frequency up conversion \cite{Kippenberg15, Kippenberg173, Diddams18}, and this presents an additional complication for device integration. In contrast, the lithium niobate microresonator demonstrated here enables simultaneous generation of coherent Kerr solitons and their second harmonics inside a single device. 
Further device development to broaden the LN comb spectrum to an octave would enable carrier envelope offset frequency measurement for f-to-2f self-referencing by simple detection of the output from a single LN resonator.


Moreover, LN exhibits a variety of outstanding properties that are fairly unique among optical media \cite{Gaylord85}. For example, in addition to the significant Kerr, quadratic, and photorefractive optical nonlinearities that are utilized here, LN exhibits a strong electro-optic Pockels effect ideal for high-speed lossless spatiotemporal modulation of nanophotonic devices. LN also exhibits a significant piezoelectric effect, about one order of magnitude larger than for aluminum nitride and gallium arsenide, which is ideal for electro-mechanical tuning of integrated photonic circuits. LN also exhibits significant pyroelectricity that can be applied for on-chip temperature sensing. Moreover, LN can be conveniently doped with rare-earth ions (\emph{e.g.},  Er$^{3+}$ and Tm$^{3+}$, etc.) \cite{Sohler05} to provide optical gain directly on chip. These important features would offer versatile approaches for controlling, manipulating, and modulating the soliton microcombs demonstrated here. 
Therefore, more generally, the demonstration of self-starting mode locked soliton microcombs in this paper opens the door towards realization of a multi-functional high-speed photonic signal processor for metrology, frequency synthesis/division, information coding, optical-optical/electro-optical waveband conversion, and other functions, all directly integrated on a single chip.

\section*{Methods}

\textbf{Device fabrication:} 
The device was fabricated on a 600-nm-thick congruent LN thin film sitting on a 3 ${\rm \mu m}$-thick buried silicon oxide layer. The ZEP-520A resist was used as a mask to define the device structure, which was patterned by electron-beam lithography. After the device pattern was defined, the LN layer was etched down by about 410~nm, via an Ar-ion milling process, forming a waveguide cross section as schematically shown in Fig.~\ref{Fig1}(b). Finally, the facet of the chip was polished to obtain good fiber-to-chip light coupling. Detailed device dimensions are provided in the Supplementary Information. 

\textbf{Dispersion characterization}
To characterize the GVD of the device, we measured the resonance frequency $\omega_\mu$ of the soliton forming mode family as a function of relative mode number ${\rm \mu}$, referred to a reference resonance at $\omega_0$ ($\mu=0$ corresponds to a cavity mode at a wavelength around 1558.7~nm). The cavity resonance frequency $\omega_\mu$ is described by a Taylor series around a reference resonance $\omega_0$ \cite{Vahala15}, $\omega_\mu = \omega_0 + \mu D_1 + \frac{1}{2} \mu^2 D_2 + \frac{1}{6} \mu^3 D_3 + \cdots$, where $D_1/(2\pi)$ is the FSR around $\omega_0$ and $D_2$ is related to the GVD ${\beta_2}$ as $D_2 = -\frac{c}{n} D_1^2 \beta_2$. Figure 1d of the main text shows $D_{\rm int} \equiv \omega_\mu -\omega_0 - \mu D_1$. Fitting the experimental data, we obtained $D_1/(2\pi)$ = 199.7~GHz and $D_2/(2\pi)$ = 1.76~MHz.

\textbf{Data availability.} The data that support this study are available from the authors on reasonable request.

\section*{Acknowledgments}
This project was supported in part by the Defense Threat Reduction Agency-Joint Science and Technology Office for Chemical and Biological Defense (grant No.~HDTRA11810047), and by the National Science Foundation under grants No.~ECCS-1810169 and ECCS-1610674. This work was performed in part at the Cornell NanoScale Facility, a member of the National Nanotechnology Coordinated Infrastructure (National Science Foundation, ECCS-1542081), and at the Cornell Center for Materials Research (National Science Foundation, DMR-1719875).

The project or effort depicted was or is sponsored by the Department of the Defense, Defense Threat Reduction Agency. The content of the information does not necessarily reflect the position or the policy of the federal government, and no official endorsement should be inferred.

\section*{Author contributions}
Y.H. designed and fabricated the sample. Y.H. and Q.-F.Y. designed and performed the experiments. J.L. did the numerical simulations. Y.H., Q.-F.Y., and J.L. analyzed  the data. J.L., R.L., B.S., and H.W. assisted in the experiments. R.L., H.L., and M.L. assisted in device fabrication. Y.H., Q.-F.Y., J.L., K.V., and Q.L. wrote the manuscript. K.V. and Q.L. supervised the project. Q.L. conceived the concept. 

\section*{Additional information}
Correspondence and requests for materials should be addressed to Q.L.

\textbf{Competing financial interests:} The authors declare no competing financial interests.

\bibliography{References}

\end{document}



	\title{Supplementary Information} 

	\author {Yang He}
	\thanks{These two authors contributed equally.}
	\affiliation{Department of Electrical and Computer Engineering, University of Rochester, Rochester, NY 14627}

	\author{Qi-Fan Yang}
	\thanks{These two authors contributed equally.}
	\affiliation{T.J. Watson Laboratory of Applied Physics, California Institute of Technology, Pasadena, California 91125, USA}

	\author{Jingwei Ling}
	\affiliation{Institute of Optics, University of Rochester, Rochester, NY 14627}
	
	\author{Rui Luo}
	\affiliation{Institute of Optics, University of Rochester, Rochester, NY 14627}
	
	\author{Hanxiao Liang}
	\affiliation{Department of Electrical and Computer Engineering, University of Rochester, Rochester, NY 14627}
	
	\author{Mingxiao Li}
	\affiliation{Department of Electrical and Computer Engineering, University of Rochester, Rochester, NY 14627}
	
	\author{Boqiang Shen}
	\affiliation{T.J. Watson Laboratory of Applied Physics, California Institute of Technology, Pasadena, California 91125, USA}
	
	\author{Heming Wang}
	\affiliation{T.J. Watson Laboratory of Applied Physics, California Institute of Technology, Pasadena, California 91125, USA}
	
	\author{Kerry Vahala}
	\affiliation{T.J. Watson Laboratory of Applied Physics, California Institute of Technology, Pasadena, California 91125, USA}
	
	\author{Qiang Lin}
	\email[Electronic mail: ]{qiang.lin@rochester.edu}
	\affiliation{Department of Electrical and Computer Engineering, University of Rochester, Rochester, NY 14627}
	\affiliation{Institute of Optics, University of Rochester, Rochester, NY 14627}

	\begin{abstract}
		
		In this supplement detailed information is provided on the following: the device design, the experimental setup, the numeric modeling of soliton comb generation with analysis of self-starting mode locking, and the characterization of key device parameters. 
		
	\end{abstract}
	
	\maketitle 
	
	\section{Device design and properties}
	
	\begin{figure*}[b!]
		\centering\includegraphics[width=1\columnwidth]{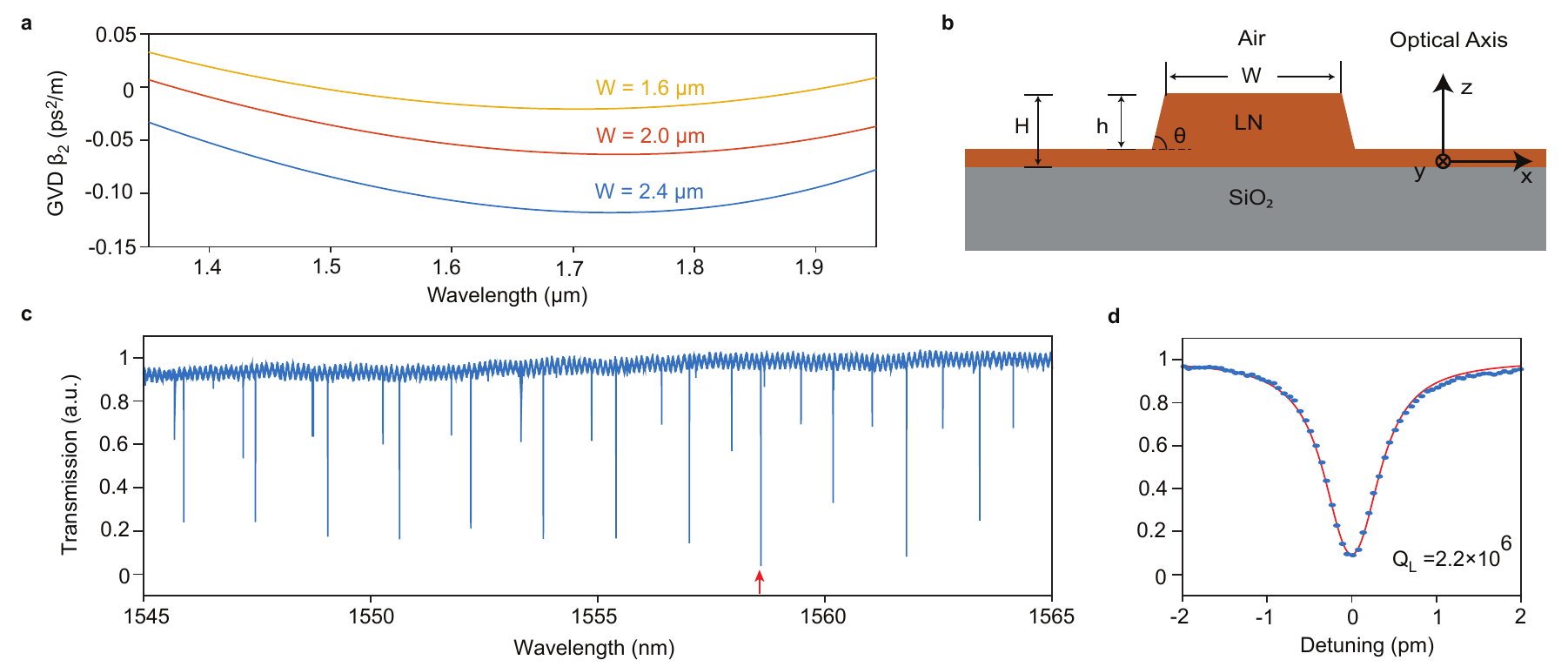}
		\caption{\footnotesize {\bf (a)} Simulated GVD curves for the quasi-TE mode of a straight waveguide with different waveguide widths (W). {\bf (b)} Schematic of the waveguide cross section, with parameters: H = 600~nm, $h=410$~nm, and $\theta=75^\circ$. {\bf (c)} Normalized transmission spectrum of the device. {\bf (d)} High resolution transmission spectrum of a quasi-TE mode, indicated as red arrow in ({\bf c}). The experiment data are shown as blue dots and the fitting curve is shown in red.}  \label{Fig.S1}
	\end{figure*}
	Control of group-velocity dispersion (GVD) in the anomalous dispersion regime is crucial for generating  soliton Kerr frequency combs. Waveguide design is possible by numerical modeling of waveguide cross sections using a finite element method solver (COMSOL) . Figure \ref{Fig.S1}(a) shows the simulated GVD curve for different waveguide widths where a diagram of the waveguide cross section is provided in  Fig.~\ref{Fig.S1}(b). Other waveguide dimensions are provided in the figure caption. A waveguide width of 2.0~${\rm \mu m}$ was chosen because it provides anomalous GVD with a reasonably small magnitude over a broad spectral band, but not so small as to be sensitive to fabrication error.  As shown in Fig.~1(d) of the main text, the GVD of the fabricated device is close to prediction from modeling. Based on the simulation, the effective refractive index $n_0$ of the device is determined to be around 2.0 at 1555~nm.  The microring is side coupled to a pulley waveguide with a width of 1.4~${\rm \mu m}$ and with a gap of 0.7~${\rm \mu m}$ as shown in Fig.~1a of the main text. 
	
	Figure \ref{Fig.S1}(c) shows the transmission spectrum of the device for coupling to the quasi-TE polarization (where the electric field of optical modes predominantly lies in the device plane.).  Fig.~\ref{Fig.S1}(d) is a higher resolution scan of a cavity mode (indicated by the red arrow in Fig.~\ref{Fig.S1}(c)) and shows that the device exhibits a high optical Q of $2.2\times 10^6$ for the loaded cavity.

	\section{Experiment setup}
	
	\begin{figure*}[b!]
		\centering\includegraphics[width=1\columnwidth]{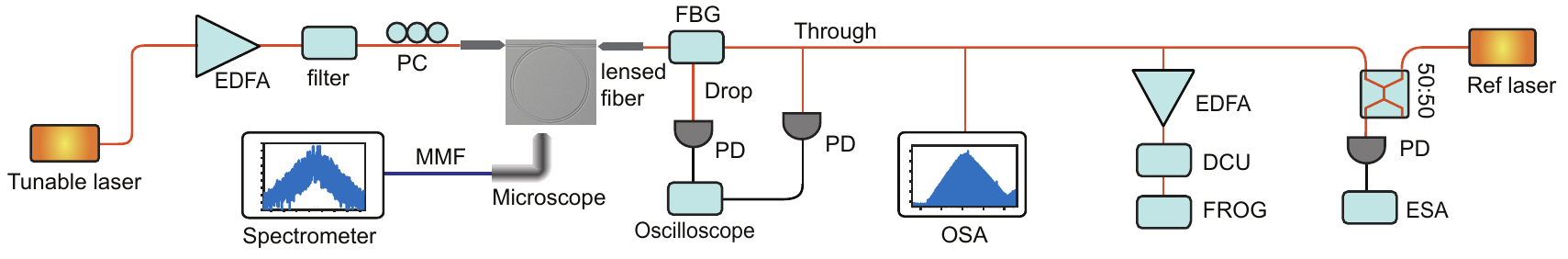}.
		\caption{\footnotesize Schematic of the experimental testing setup.}  \label{Fig.Exp}
	\end{figure*}
	Figure \ref{Fig.Exp} shows the schematic of the experimental testing setup. A continuous-wave (CW) telecom-band tunable laser was used as the pump laser. The laser power was boosted by an erbium-doped fiber amplifier (EDFA) whose amplified spontaneous emission (ASE) noise was suppressed by a tunable optical filter. The polarization of the laser was adjusted by a polarization controller (PC) to match the quasi-TE polarization of the cavity mode. The pump light was coupled to the device chip through a lensed fiber. The comb light output from the device chip was collected by another lensed fiber. A fiber Bragg grating (FBG) was used to separate the pump light from the frequency comb, and the pump light (directed to the drop port in the schematic) was photodetected (PD) to monitor the cavity transmission (Fig.~1{\bf e,f} in the main text). The comb light transmitted to the through port was measured using a photodetector (Fig.~1{\bf g,h} in the main text), an optical spectrum analyzer (OSA), and a frequency-resolved optical gating (FROG) for soliton temporal waveform characterization. In order to reach the minimum working power of FROG, an EDFA boosted the comb power and a dispersion compensation unit (DCU) was used to compensate the dispersion introduced by the EDFA. As the repetition rate of the device is too large to be measured by an electrical spectrum analyzer (ESA), the coherence of individual comb lines was studied by heterodyne with a reference CW laser and measurement of the electrical spectrum by an ESA. The generated second harmonic of the Kerr soliton was characterized from the light scattered from the device into free space. The scattered light was collected by an imaging microscope whose output was either recorded by a CMOS camera (spectral response: 400-1000~nm) or was delivered, via a multimode fiber (MMF), to a spectrometer that covers the near infrared spectral band around 700-900~nm.

	\section{On the theoretical modeling of self-starting soliton mode locking}
	
	\subsection{The Lugiator-Lefever equation}
	
	The generation of a Kerr frequency comb inside a microresonator is well described by the Lugiato-Lefever equation (LLE) \cite{Lugiato87, Coen132}
	\begin{equation}
	t_R\frac{\partial A(t,\tau)}{\partial t}=\Big[-\alpha-i\delta_0+iL\sum_{k\geq 2}\frac{\beta_k}{k!}(i\frac{\partial}{\partial \tau})^k+ i\gamma L|A|^2\Big]A+\sqrt{\theta}A_{in}.
	\label{LLE}
	\end{equation}
	where $A$ is the optical field normalized such that $|A|^2$ is normalized to optical power.  Also,  $t_R$ is the roundtrip time, $L$ is the cavity length, $\alpha=(\alpha_0+\theta)/2$ is the total loss (where $\alpha_0$ is the intrinsic absorption coefficient and $\theta$ is the transmission coefficient determined by the coupling between the resonator and the coupling waveguide),  $\beta_k$ is the $k^{\rm th}$-order dispersion coefficient,  $\gamma = \frac{n_2\omega_0}{cA_{\rm eff}}$ is the nonlinear parameter (where $n_2$ is Kerr nonlinear coefficient and $A_{\rm eff}$ is the effective mode area), and $\delta_0=2\pi l-\phi_0$ is the phase detuning between the $l^{\rm th}$-order reference resonance mode and the driving field. In Eq.~(\ref{LLE}), $\tau$ is the \emph{fast} time variable used to describe the detailed dynamics of comb within a round-trip time, and $t$ is the \emph{slow} time variable used to describe the comb evolution over a time scale longer than the round-trip time \cite{Coen132}.
	
	Considering the periodic condition of a ring resonator for the intra-cavity field, the field $A^{(m+1)}(0,\tau)$ at the beginning of the next roundtrip will be the summation of input filed $A_{in}$ and the original filed $A^{(m)}(L,\tau)$ after a roundtrip length $L$ \cite{Coen132}:
	\begin{equation}
	A^{(m+1)}(0,\tau)=\sqrt{\theta}A_{in}+\sqrt{1-\theta}A^{(m)}(L,\tau)e^{i\phi_0}.
	\label{Resonator}
	\end{equation}
	
	\noindent In general, the properties of a microresonator are well described by device parameters such as optical Q, photon decay rate $\Gamma_t$ (related to optical Q as $\Gamma_t = \frac{\omega_0}{Q}$), and the laser-cavity frequency detuning $\Delta_0 = \omega_p - \omega_0$ (where $\omega_p$ is the pump frequency and $\omega_0$ is the frequency of the passive cavity mode being pumped and in the absence of the optical power). Therefore, Equation (\ref{LLE}) can be written in a more intuitive way as 
	\begin{equation}
	\frac{\partial A(t,\tau)}{\partial t} = \Big[-\frac{\Gamma_t}{2} + i \Delta_0 + iv_g \sum_{k\geq 2}\frac{\beta_k}{k!}(i\frac{\partial}{\partial \tau})^k+ i\gamma v_g |A|^2\Big]A + \sqrt{\frac{\theta}{t_R^2}}A_{in},
	\label{LLE2}
	\end{equation}
	where $v_g = \frac{L}{t_R}$ is the group velocity, $\Delta_0$ is related to $\delta_0$ as $\Delta_0 = -\frac{\delta_0}{t_R}$, and $\Gamma_t$ is related to $\alpha$ as $\Gamma_t = 2 \alpha v_g$. The final soliton state will be determined by the detuning $\Delta_0$, the dispersion $\beta_k$, Kerr nonlinearity $n_2$, and the optical Q of the loaded cavity ($Q=\frac{\omega_0}{\Gamma_t}$) \cite{Coen13}.
	
	\subsection{Description and incorporation of the photorefractive effect}
	
	Equation (\ref{LLE2}) (or Equation (\ref{LLE})) includes only the optical Kerr effect and the dispersion effect. It is not adequate to describe our device where other nonlinear effects are expected to play important roles. A dominant one is the photorefractive effect \cite{GunterBook}, where the refractive index of the device material decreases with increased optical power inside the resonator, leading to a blue shift of the cavity resonance. The photorefractive effect is essentially an electro-optic effect induced by the space charge electric field produced by photo-excitation (to certain defects) \cite{GunterBook}. Therefore, the photorefractive-induced resonance tuning can be described as $\delta \omega_0 = g_E E_{sp}$, where $g_E$ represents the electro-optic coupling coefficient and $E_{sp}$ represents the induced space-charge electric field \cite{Sun17}. 
	
	As shown below, the photorefractive effect responds very slowly to changes in cavity optical power in comparison to both the roundtrip time ($\sim$5~ps) and the photon lifetime ($\sim$1.8~ns) of the resonator.  Therefore, the photorefractive-induced resonance tuning can be described in an adiabatic fashion.  As such, the laser-cavity detuning term in Eq.~(\ref{LLE2}), $\Delta_0$, is replaced by $\Delta = \Delta_0 - \delta \omega_0 = \Delta_0 - g_E E_{sp}$, and Eq.~(\ref{LLE2}) now becomes
	\begin{equation}
	\frac{\partial A(t,\tau)}{\partial t} = \Big[-\frac{\Gamma_t}{2} + i \Delta_0 - i g_{E} E_{sp} + iv_g \sum_{k\geq 2}\frac{\beta_k}{k!}(i\frac{\partial}{\partial \tau})^k+ i\gamma v_g |A|^2\Big]A + \sqrt{\frac{\theta}{t_R^2}}A_{in}.
	\label{LLE3}
	\end{equation}Furthermore, the dynamics of the space-charge electric field can be described by a simple excitation-relaxation process \cite{Sun17}. 
	\begin{equation}
	\frac{d(E_{sp})}{dt}=-\Gamma_{sp} E_{sp}+\eta_{sp}\overline{|A|^2},
	\label{Dyna}
	\end{equation}where $\Gamma_{sp}$ is the relaxation rate of the space charge field and $\eta_{sp}$ is the optical generation coefficient. Also, $\overline{|A|^2} = \frac{1}{t_R} \int_0^{t_R} {|A|^2 d\tau} $ is the round trip average power. Equation (\ref{LLE3}) and (\ref{Dyna}) form a complete set of equations describing Kerr comb generation under the influence of the photorefractive effect. We have neglected the thermo-optic effect since, as discussed in the main text, its effect is much smaller compared with the photorefractive effect. Numerically,  Eq.~(\ref{LLE3}) is solved by the split-step Fourier method, where the time-dependent space-charge field is obtained from Eq.~(\ref{Dyna}) as  
	\begin{eqnarray}
	E_{sp}^{(m+1)} = (1- \delta t \Gamma_{sp}) E_{sp}^{(m)} + \delta t \eta_{sp} \overline{|A|^2}, \label{Dyna2}
	\end{eqnarray}
	since the time step $\delta t$ used for the split-step Fourier method is always much smaller than the photorefractive relaxation time $\frac{1}{\Gamma_{sp}}$.
	
	In addition to the photorefractive effect described above, lithium niobate exhibits a pyroelectric effect \cite{Gaylord85}. Specifically, when an optical wave is launched into the resonator, photothermal heating (say, via material absorption) will increase the device temperature which in turn produces a material polarization and a space charge field that will blue shift the cavity resonance via the electro-optic effect.  However, as the pyroelectricity-induced space charge field has a magnitude that is linearly proportional to the induced temperature increase, which in turn depends linearly on the optical power, Equation (\ref{Dyna}) can still apply (although via an immediate temperature change). Therefore, while the time constants are slightly different, the photorefractive effect and the pyroelectric effect have a similar functional impact on the cavity resonance. For simplicity, we use Eq.~(\ref{Dyna}) to describe the overall effect, lumping them together as the ``photorefractive effect".

	\subsection{Characterization of key parameters}
	
	Numeric modeling via Eqs.~(\ref{LLE3}) and (\ref{Dyna2}) requires detailed information about the device parameters. The linear properties, such as cavity Q, external coupling, and GVD, can be obtained from linear characterization of the device as described in Section A and in Fig.~1 of the main text. The cavity length and the effective mode area are obtained from the finite element simulation of the device.  The $n_2$ of lithium niobate was measured directly in our device via the threshold power of optical parametric oscillation, which is given by \cite{Vahala15}:
	\begin{equation}
	P_{th}=\frac{\pi n\omega_0 A_{\rm eff}}{4\eta n_2}\frac{1}{D_1Q^2}.
	\end{equation}The threshold power was determined by decreasing the on-chip pump power while scanning the wavelength quickly from blue to red until the onset of the parametric oscillation. The threshold power was thus measured to be 4.2~mW, and we determined the Kerr nonlinear coefficient to be $n_2 = 1.8\times10^{-19}~{\rm m^2/W}$.
	
\begin{figure*}[b!]
		\centering\includegraphics[width=\columnwidth]{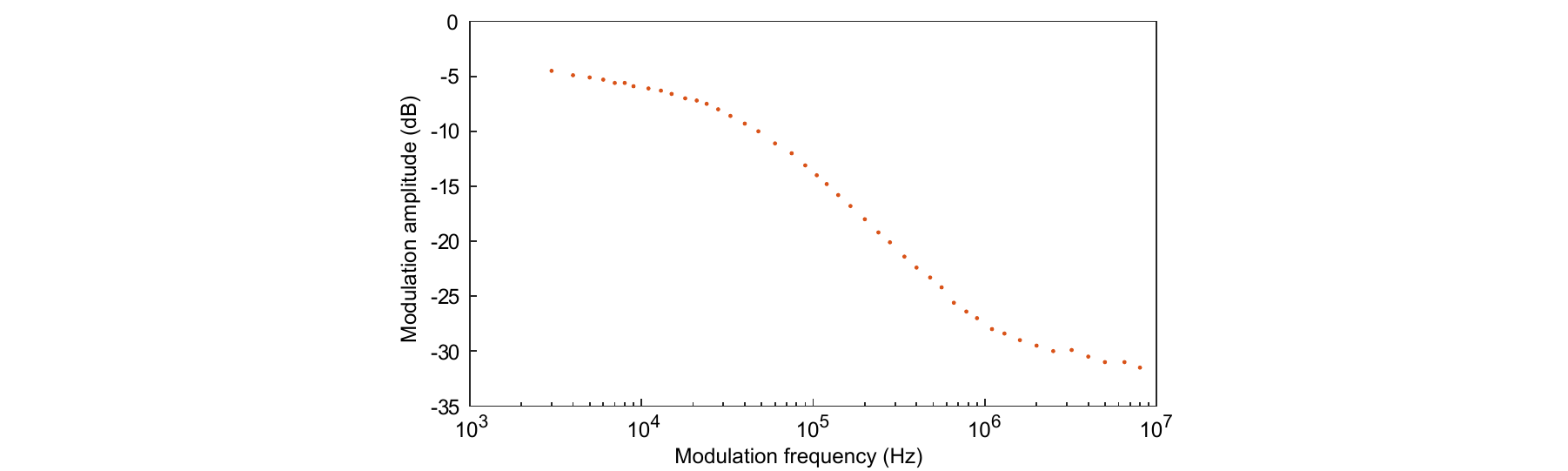}.
		\caption{\footnotesize The frequency response of the photorefractive effect.} \label{photo_ref}
	\end{figure*}

	To determine the photorefractive response time constant,  we employed a pump-probe approach \cite{Vahala05, Lu14} where a sinusoidal modulation on a pump wave launched into a cavity mode modulates the refractive index which, in turn, is sensed by a weak probe wave launched into a separate cavity mode. The detailed experimental setup can be found in Ref.~\cite{Lu14}. Figure \ref{photo_ref} shows the recorded modulation response spectrum. The photorefractive effect exhibits a non-Lorentzian spectral response shape, indicating a potentially more complex (multi-time-constant) relaxation process, as is expected \cite{Sun17}. The modulation response exhibits a 3-dB bandwidth of about 20kHz, corresponding to a dominant relaxation time constant of $\sim$8~${\rm \mu s}$. For simplicity, we used this time response to describe the photorefractive dynamics as given in Eq.~(\ref{Dyna}).
	
	Analysis of Eqs.~(\ref{Dyna}) and (\ref{LLE3}) shows that the essential parameter governing the strength of photorefraction-induced cavity resonance tuning is $\rho_{pr} \equiv \frac{g_{E} \eta_{sp}}{\Gamma_{sp}}$. However, characterization of this parameter turns out to be quite complicated, since the field generation coefficient $\eta_{sp}$ is power dependent \cite{Sun17}. This is particularly challenging during the Kerr comb generation since the intracavity field dynamics affect the overall intracavity energy (as shown in the next section). As such, we treat $\rho_{pr}$ as a fitting parameter. Detailed modeling and analysis shows that a value of $\rho_{pr} = 8\times 10^{-3}~{\rm THz/W}$ provides a reasonably good description of the system, and this value is used in the numerical modeling of comb generation. 
	
	\subsection{Numerical modeling of soliton comb generation, power spikes and substructure}
\begin{figure*}[htbp]
		\centering\includegraphics[width=1\columnwidth]{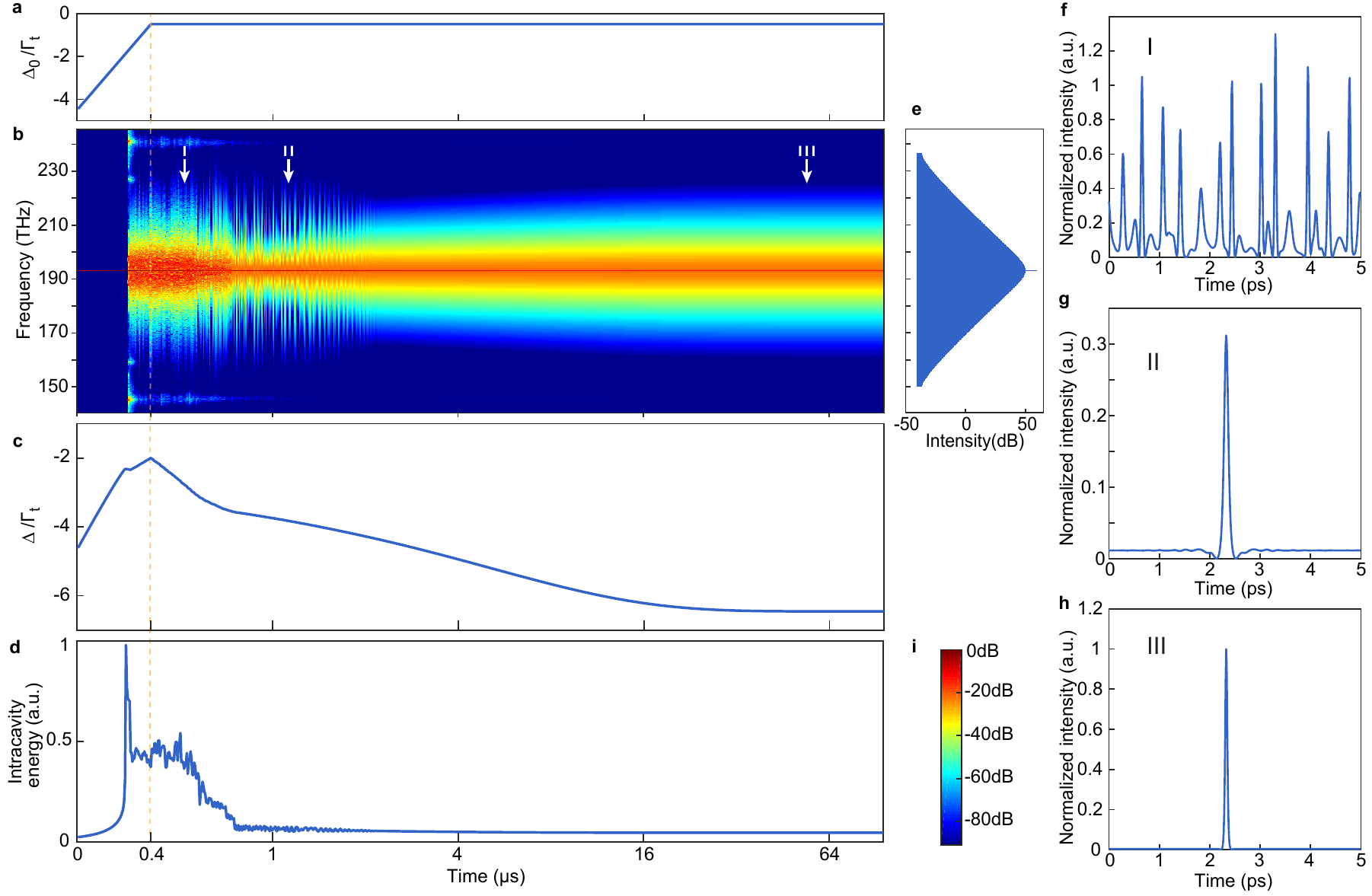}
		\caption{\footnotesize {\bf Numerical modeling of the self-starting mode locking process.} {\bf a.} Time dependence of the input pump frequency. The pump frequency is tuned linearly within 0.4~${\rm \mu s}$ to a value of a value of $\omega_p - \omega_0(t_0) = - \frac{\Gamma_t}{2} $ ($t_0 = 0.4~{\rm \mu s}$) and remains constant afterwards, where $\Gamma_t$ is the linewidth of the loaded cavity and $\omega_0(0)$ is the resonance frequency of the passive cavity in the absence of the input power. The pump power is fixed at 33~mW. {\bf b.} Simulated comb spectrum as a function of time within 100~${\rm \mu s}$. The corresponding color bar is shown in ({\bf i}). {\bf d.} Laser-cavity detuning, $\Delta \equiv \omega_p - \omega_0(t)$ (normalized by $\Gamma_t$), as a function of time where $\omega_0(t)$ is the time-dependent cavity resonance under the photorefractive effect. {\bf d.} Intracavity energy as a function of time. Note that in ({\bf a})-({\bf d}) the horizontal time scale is set to be a linear scale within the initial 0-0.4~${\rm \mu s}$ so as to show the detailed tuning of pump frequency and increase of intracavity energy, but to be a log scale within (0.4-100)~${\rm \mu s}$ to show the detailed evolution of the mode locking process. {\bf e.} Simulated spectrum of the mode-locked single soliton state at 100~${\rm \mu s}$ showing a ${\rm sech^2}$ spectral shape. {\bf f-h.} Temporal waveforms of the intracavity optical field at three different times indicated by the arrow I-III in ({\bf b}). The intensities of three figures are all normalized to the peak intensity of the single soliton in ({\bf h}). {\bf i.} Color bar for ({\bf b}). }\label{Fig_FreqTuning_1}
	\end{figure*}

	With the key parameters given above, it is possible to use Eqs.~(\ref{LLE3}) and (\ref{Dyna2}) to model the soliton comb generation via  by the split-step Fourier method. 
	A first example is meant to simulate the conditions in Fig 2 of the main text wherein the pumping laser is tuned towards the cavity resonance from the red detuned side. This process was discussed in some detail in the main text and leads to power spikes followed by stable soliton generaion. Moreover, the power spikes consist of an unstable initiation followed potentially by a rapid cascade of higher to lower number soliton states. To simulate this process the pump power is set at 33~mW and the pump frequency is quickly tuned to a position that is one-half cavity-linewidth red detuned from the cold cavity resonance (Fig.~\ref{Fig_FreqTuning_1}a).  Figure \ref{Fig_FreqTuning_1} shows the resulting simulated evolution of the comb generation process. When the pump frequency is tuned into the cavity resonance the intracavity optical energy initially grows (Fig.~\ref{Fig_FreqTuning_1}d) until the parametric oscillation threshold is attained. The oscillation quickly evolves into a broad spectral band via cascaded four-wave mixing (Fig.~\ref{Fig_FreqTuning_1}b), which creates a random waveform in the time domain (Fig.~\ref{Fig_FreqTuning_1}f). At the same time, the intracavity optical energy induces a photorefractive response that shifts the cavity resonance gradually towards the blue, leading to increased laser-cavity detuning ($|\Delta |$ increases) (Fig.~\ref{Fig_FreqTuning_1}c). Around a time of $t_1 = 1.2~{\rm \mu s}$, the laser-cavity detuning has shifted to a value ($\Delta(t_1) \equiv \omega_p - \omega_0(t_1) = -3.6~\Gamma_t $) that is large enough to allow the comb to transit into a breather soliton state (Fig.~\ref{Fig_FreqTuning_1}b,g). This results in a rapid decrease of the intracavity energy (Fig.~\ref{Fig_FreqTuning_1}d).  However, the photorefractive response is still strong enough to continuously blue shift the cavity resonance. The resulting gradual increase of cavity-laser detuning causes the soliton spectrum to broaden by a certain extent. Eventually, around $t_2 = 41~{\rm \mu s}$, the cavity resonance settles to a stable value (Fig.~\ref{Fig_FreqTuning_1}c), resulting in a stable mode locked single-soliton state (Fig.~\ref{Fig_FreqTuning_1}h). 
	
\begin{figure*}[htbp]
		\centering\includegraphics[width=1\columnwidth]{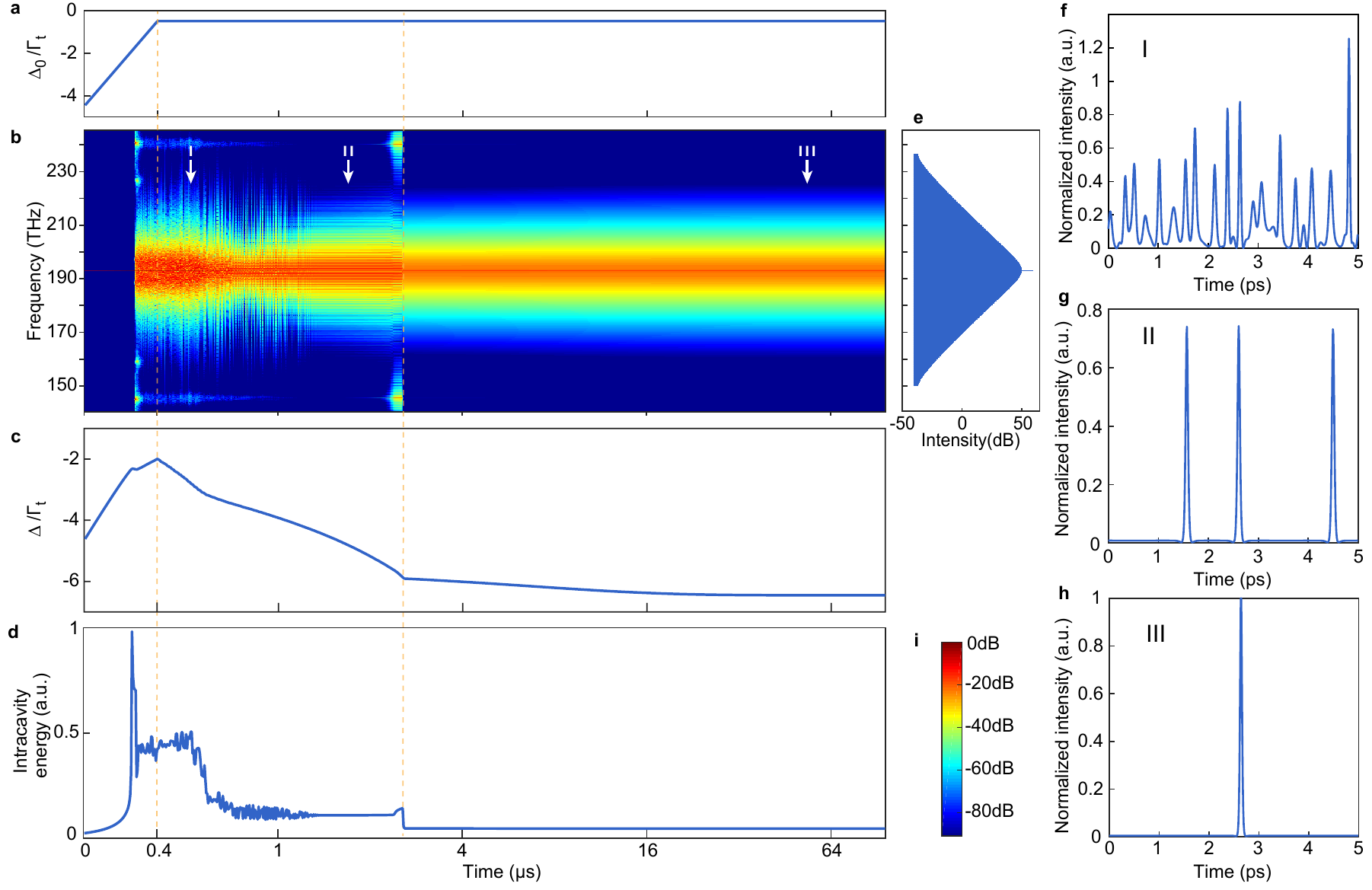}
		\caption{\footnotesize {\bf Numerical modeling of the self-starting mode locking process.} Same as Fig.~\ref{Fig_FreqTuning_2} but showing an intermediate multi-soliton state. }\label{Fig_FreqTuning_2}
	\end{figure*}

	Figure \ref{Fig_FreqTuning_1} describes the condition in which a power spike (see main text discussion) evolves immediately into a single soliton state. However, as observed in Fig. 2 of the main text, there can also be meta-stable soliton states within the spikes. Importantly, these states are also observed in the numerical modeling. This is shown in the example of Fig.~\ref{Fig_FreqTuning_2}, where the photorefractive effect causes the Kerr comb to transit first into a three-soliton state that exists over a short time period of 1.6~${\rm \mu s}$ before eventually transiting into a stable single soliton. 
	
	\section{Discussion}
	
	The analysis in the previous sections reveals the underlying mechanism of the self-starting process of soliton mode locking in our device and the essential role played by the photorefractive effect. However, it should be noted that it is based upon a simplified single-relaxation model of the photorefractive effect (Eq.~(\ref{Dyna})) with a constant field generation coefficient. As shown in Fig.~\ref{photo_ref}, the frequency response of photorefraction is fairly complicated with potential multi-scale time responses \cite{GunterBook}. Moreover, the field generation coefficient could be power dependent \cite{Sun17}. All these factors might impact the exact behavior of the self-starting soliton mode locking and requires further exploration in the future.

	\bibliography{supplement}
	
	\if{
		
	}\fi
